\documentclass[aip,reprint,amsmath,amssymb,superscriptaddress]{revtex4-2}

\usepackage{graphicx}
\usepackage{dcolumn}
\usepackage{bm}
\usepackage{color}

\begin{document}

\title{The effect of finite mass in cavity-QED calculations}

\author{Ankita Nair}
\affiliation{Department of Physics and Astronomy, Vanderbilt
University, Nashville, Tennessee, 37235, USA}

\author{Vikas Bharti}
\affiliation{Department of Physics and Astronomy, Vanderbilt
University, Nashville, Tennessee, 37235, USA}

\author{Yetmgeta S. Aklilu}
\affiliation{Department of Physics and Astronomy, Vanderbilt
University, Nashville, Tennessee, 37235, USA}

\author{K\'alm\'an Varga}
\email{kalman.varga@vanderbilt.edu}
\affiliation{Department of Physics and Astronomy, Vanderbilt
University, Nashville, Tennessee, 37235, USA}

\begin{abstract}
The effect of finite nuclear mass is investigated in coupled light
matter systems  in cavity quantum electrodynamics (cavity QED)
using the Pauli-Fierz Hamiltonian. Three
different systems, the He atom, the H$^-$ ion and the H$_2^+$ ion is
investigated. There are small, but significant differences in the
behavior of the binding energies as the function of the coupling
strength. The probability of coupling to light is found to be very small
but even this small coupling has a very strong effect on the energies
of the systems.
\end{abstract}
\maketitle

\section{Introduction}
Cavity quantum electrodynamics  is a powerful platform for
implementing quantum sensors  
\cite{degen_quantum_2017,reiserer_nondestructive_2013}, 
memories\cite{specht_single-atom_2011,ma_high-performance_2022}, 
and networks \cite{reiserer_cavity-based_2015,daiss_quantum-logic_2021,
boozer_reversible_2007,
ritter_elementary_2012,briegel_quantum_1998,muralidharan_optimal_2016}. 
In cavity QED, a
quantum emitter, such as an atom, molecule or a quantum dot, is coupled to the
electromagnetic modes confined within a cavity. Strong interactions between 
cavity photons and molecular systems can result in the formation of
hybrid light-matter states called
polaritons. These polaritons can exhibit significantly different
chemical and physical properties compared to their individual components.
The strongly coupled light-matter states can dramatically change
physical and chemical processes. For example, a cavity can enhance the
energy transport in molecules \cite{sandik_cavity-enhanced_2024}, 
suppress photochemical reactions \cite{galego_suppressing_2016}, or
induce catalytic processes
\cite{campos-gonzalez-angulo_resonant_2019,lather_cavity_2019}. 
Cavities can also change reactivity
\cite{galego_cavity_2019,thomas_tilting_2019}, photoisomerisation
\cite{fregoni_manipulating_2018}, ionization
\cite{li_collective_2021}, 
excited states \cite{stranius_selective_2018} or electron captures processes
\cite{cederbaum_making_2024}.

The ability to manipulate the physical and chemical characteristics of
materials through interaction with light has sparked significant
experimental \cite{https://doi.org/10.1002/anie.201107033,
Balili1007,PhysRevLett.114.196403,Xiang665,
doi:10.1021/acsphotonics.0c01224,Coles2014,Kasprzak2006,PhysRevLett.106.196405,Plumhof2014,
https://doi.org/10.1002/adma.201203682,Wang2021,BasovAsenjoGarciaSchuckZhuRubio+2021+549+577}
and theoretical \cite{PhysRevLett.114.196402,PhysRevLett.121.253001,
PhysRevLett.128.156402,Riso2022,
https://doi.org/10.1002/qua.26750,
PhysRevLett.122.017401,DiStefano2019,PhysRevX.5.041022,
PhysRevLett.116.238301,Galego2016,Shalabney2015,doi:10.1021/acsphotonics.9b00648,Schafer4883,
Ruggenthaler2018,Flick15285,Flick3026,PhysRevLett.123.083201,Mandal,doi:10.1021/acsphotonics.9b00768,
doi:10.1021/acs.nanolett.9b00183,FlickRiveraNarang,PhysRevLett.121.113002,Garcia-Vidaleabd0336,
Thomas615,PhysRevResearch.2.023262,doi:10.1063/5.0036283,doi:10.1063/5.0038748,doi:10.1063/5.0039256,
doi:10.1063/5.0012723,doi:10.1063/5.0021033,acs.jpcb.0c03227,PhysRevLett.119.136001,
doi:10.1063/5.0012723,Flick15285,doi:10.1021/acsphotonics.7b01279,PhysRevA.98.043801,
doi:10.1021/acs.jpclett.0c01556,doi:10.1021/acs.jctc.0c00618,doi:10.1021/acs.jpclett.0c03436,
doi:10.1021/acs.jctc.0c00469,PhysRevB.98.235123,PhysRevLett.122.193603,
Szidarovszky_2020,doi:10.1021/acs.jpclett.1c01570,doi:10.1021/jacs.2c00921,
doi:10.1063/5.0095552,doi:10.1021/acs.jpclett.1c02659,
doi:10.1021/jacs.1c13201,Cederbaum2021,
10.1063/5.0142403,doi:10.1021/acs.jctc.4c00763,
10.1063/5.0216993,cui24} interest.
Several excellent review articles have been published, highlighting
the current state of experimental and theoretical approaches related
to light-matter interactions in cavities.  These include reviews about
the  properties of hybrid light-matter states
\cite{doi:10.1021/acs.accounts.6b00295,doi:10.1063/PT.3.4749}, ab
initio calculations \cite{Ruggenthaler2018,doi:10.1063/5.0094956} and
molecular polaritonics
\cite{doi:10.1146/annurev-physchem-090519-042621,doi:10.1021/acsphotonics.2c00048,
doi:10.1021/acsphotonics.1c01749}.

The theoretical and computational description of the coupled
light-matter system is challenging because the already complex
solution of the quantum many-body problem of the interaction between
electrons and nuclei is further complicated by the addition of photon
degrees of freedom. In recent years, a variety of approaches have been
proposed that go beyond the simple two-level atom model
\cite{Jaynes1962ComparisonOQ}. Most of
these approaches are based on successful many-body quantum methods
adapted to the interaction with photons. The use of the Pauli-Fierz
(PF) non-relativistic QED Hamiltonian has been found to be the most
useful framework \cite{Ruggenthaler2018,Rokaj_2018,Mandal,acs.jpcb.0c03227,PhysRevB.98.235123}
for practical calculations. The PF Hamiltonian
is a sum of electronic and photonic Hamiltonians, along with a
cross-term describing the electron-photon interaction.
Due to this cross term, one has to use a coupled
electron-photon wave function,
\begin{equation}
\sum_{\vec{n}} \Phi_{\vec{n}}(\mathbf{x})\chi_{\vec{n}}
\label{coupled}
\end{equation}
where $\mathbf{x}=(\mathbf{r}_1,\mathbf{r}_2,{\ldots},
\mathbf{R}_1,\mathbf{R}_2,{\ldots})$ are the spatial coordinates of
the electrons and
nuclei and $\vec{n}=(n_1,n_2,{\ldots} N_p)$ are the quantum numbers of
the photon
modes. The occupation number basis, $\chi_{\vec{n}}=\vert
n_1,_n2,{\ldots},N_p\rangle$,
is used to represent the bosonic Fock-space of photon
modes. 

Wave function based approaches
\cite{PhysRevLett.127.273601,doi:10.1063/5.0066427,doi:10.1063/5.0078795,
doi:10.1063/5.0038748,PhysRevX.10.041043,10.1063/5.0230565,doi:10.1021/acs.jctc.3c01207,
10.1063/5.0142403,doi:10.1021/acs.jctc.4c00763,
10.1063/5.0216993,cui24}
typically use coupled electron-photon wave functions and the product
form significally increases the dimensionality. 
The coupled cluster (CC) \cite{doi:10.1063/5.0078795,doi:10.1063/5.0038748,PhysRevX.10.041043,
PhysRevResearch.2.023262} approach used in this context defines a
reference wavefunction as a direct product of a Slater determinant of
Hartree-Fock states and a zero-photon number state. The ground state
QED-CC wavefunction is then obtained by applying an exponentiated
cluster operator to this product state. The key benefit of this
approach is its systematic improbability.
Another approach, the recently introduced cavity quantum electrodynamics 
complete active
space configuration method \cite{doi:10.1021/acs.jctc.3c01207, 
10.1063/5.0230565,matousek_polaritonic_2024}
uses linear combination of determinants of electronic orbitals and
photon-number states to describe the system. Approaches using
perturbation theory are also developed
\cite{bauer_perturbation_2023,szidarovszky_efficient_2023,cui24}. 

Density based approaches such as the quantum electrodynamics density
functional theory (QEDFT)
\cite{doi:10.1063/5.0039256,doi:10.1021/acsphotonics.7b01279,
PhysRevA.98.043801,PhysRevA.90.012508,
PhysRevLett.110.233001,doi:10.1063/5.0021033,doi:10.1063/5.0057542}.
combine the very efficient density functional methods with the photon
degrees of freedom. QEDFT is an exact reformulation of the PF
Hamiltonian-based many-body wave theory. In practical QEDFT
applications, one must develop good approximations of the
fields and currents so that the auxiliary non-interacting system
generates the same physical quantities as the interacting system. An
alternative approach \cite{10.1063/5.0123909} uses a tensor product of real 
space density functional theory representation and photon-number states
bringing the QEDFT closer to the wave function based approaches.

The PF Hamiltonian can be solved numerically exactly for a one electron 
atom or ion using product
states of Gaussian basis functions and photon number states
\cite{PhysRevA.110.043119}. The properties of small atoms and molecules
can also be accurately calculated by using a product state of
correlated gaussian basis states \cite{RevModPhys.85.693} and 
photon number states \cite{PhysRevLett.127.273601,doi:10.1063/5.0066427}.

In this work, the stochastic variational method
\cite{suzuki1998stochastic,PhysRevC.52.2885,PhysRevLett.127.273601}
will be used to optimize light-matter coupled wave functions. The calculations
can reach the same accuracy as conventional high precision calculations
for small system \cite{PhysRevLett.127.273601,doi:10.1063/5.0066427}. 

The main aim of this work is to study the difference between the
Born-Oppenheimer (infinite nuclear mass) and the non Born-Oppenheimer
(finite nuclear mass) approaches. In most
calculations \cite{PhysRevLett.127.273601,doi:10.1063/5.0066427,
doi:10.1063/5.0078795,doi:10.1063/5.0038748,PhysRevX.10.041043,
10.1063/5.0142403,doi:10.1021/acs.jctc.4c00763,10.1063/5.0216993,cui24}
the nuclear masses  assumed to be infinite and the nuclei are not
treated quantum mechanically. The PF Hamiltonian, 
however, also contains a
nuclei-photon coupling term and the dipole self-interaction (DSI) depend on
the nuclear coordinates as well. This work will elucidate the role of
these terms using small molecules and ions as test cases. 

The spatial wave functions will be represented by Explicitly Correlated  
Gaussian (ECG) basis functions \cite{RevModPhys.85.693}.
The advantage of  the approach is that the matrix elements are
analytically available
\cite{suzuki1998stochastic,doi:10.1063/1.4974273,Zaklama2019} and  it
allows very accurate calculations of energies and wave functions
\cite{RevModPhys.85.693,Zhang2015,PhysRevB.93.125423,PhysRevB.61.13873,
PhysRevLett.83.5471,PhysRevB.59.5652,PhysRevB.63.205308}.

\section{Formalism}
\subsection{Hamiltonian} 
\label{A}
The Hamiltonian of the system is
\begin{equation}
    H=H_e+H_{ph}.
\end{equation}
$H_e$ is the usual electronic Coulomb Hamiltonian, and $H_{ph}$ is the electron
photon interaction. The electron-photon interaction is described by using the PF nonrelativistic QED Hamiltonian. 
The PF Hamiltonian can be derived
\cite{Ruggenthaler2018,Rokaj_2018,Mandal,acs.jpcb.0c03227,PhysRevB.98.235123} by
applying the Power-Zienau-Woolley gauge transformation \cite{Zienau}, 
with a unitary phase transformation on the minimal coupling ($p\cdot A$) Hamiltonian in the Coulomb gauge, 
\begin{eqnarray}
H_{ph}&=&{\frac{1}{2}} \sum_{\alpha=1}^{N_p} \left[
p_{\alpha}^2+\omega_\alpha^2\left(
q_{\alpha}-\frac{{\vec{\lambda}_{\alpha}}} {{\omega_\alpha}}\cdot\vec{D}\right)^2
\right]\nonumber \\
&=&H_p+H_{ep}+H_d,
\end{eqnarray}
where $\vec{D}$ is the dipole operator. The photon fields are described by quantized oscillators. 
$q_\alpha={\frac{1}{ \sqrt{2\omega_\alpha}}}(\hat{a}^+_\alpha+\hat{a}_\alpha)$ is the displacement field and 
$p_\alpha$ is the conjugate momentum.
This Hamiltonian describes $N_p$ photon modes with frequency
$\omega_{\alpha}$ and coupling  $\vec{\lambda}_{\alpha}$. 
The coupling term is usually written as \cite{PhysRevA.90.012508}
\begin{equation}
    \vec{\lambda}_{\alpha}=\sqrt{4\pi}\,S_\alpha(\vec{r})\vec{e}_\alpha,
\end{equation}
where $S_\alpha(\vec{r})$ is the mode function at position $\vec{r}$
and $\vec{e}_\alpha$ 
is the transversal polarization vector of the photon modes.
The  three components of the electron-photon 
interaction are as follows: The photonic part is
\begin{equation}
H_{p}=\sum_{\alpha=1}^{N_p}\left(\frac{1}{2} p_{\alpha}^{2}+\frac{\omega_{\alpha}^{2}}{2} q_{\alpha}^{2}\right) = 
\sum_{\alpha=1}^{N_p} \omega_{\alpha}\left(\hat{a}_{\alpha}^{+} \hat{a}_{\alpha}+\frac{1}{2}\right),
\label{phh}
\end{equation}
and the interaction term is
\begin{equation}
    H_{ep}=-\sum_{\alpha=1}^{N_p}\omega_{\alpha}q_\alpha
    \vec{\lambda}_{\alpha}\cdot\vec{D}=
    -\sum_{\alpha=1}^{N_p}\sqrt{\frac{\omega_{\alpha}}{
    2}}(\hat{a}_{\alpha}+\hat{a}_{\alpha}^+)\vec{\lambda}_{\alpha}\cdot\vec{D}.
\label{hep}
\end{equation}

The dipole self-interaction is defined as
\begin{equation}
H_{d}={\frac{1}{2}} \sum_{\alpha=1}^{N_p} \left(\vec{\lambda_{\alpha}}
\cdot \vec{D}\right)^{2},
\label{dsh}
\end{equation}
and the  importance of this term for the existence of a ground state is discussed in Ref. \cite{Rokaj_2018}.

In the following, we will assume that there is only one important
photon mode with frequency $\omega$ and coupling $\vec{\lambda}$. Thus the suffix $\alpha$ is omitted in what follows. The
formalism can be easily extended for many photon modes but here we
concentrate on calculating the matrix elements and it is sufficient to
use a single-mode. 

For one photon mode Eqs. \eqref{phh}, \eqref{hep}, 
and \eqref{dsh} can be simplified and the Hamiltonian becomes
\begin{equation}
    H=T+V+U+\omega\left(\hat{a}^+\hat{a}+{1\over 2}\right)+\omega\vec{\lambda}\cdot\vec{D}q+{\frac{1}{2} }(\vec{\lambda}\cdot\vec{D})^2,
    \label{Hamil}
\end{equation}
In the following we assume that the system has $N$ particles with
position $\vec{r}_i$, mass $m_i$ and charge $q_i$. The position of the
$N_{nuc}$
particles with infinite mass will be fixed at $\vec{R}_i$.
The kinetic operator is 
\begin{equation}
    T=-\frac{1}{2}
    \sum_{i=1}^{N}\frac{1}{m_i}\vec{\nabla}^2_{\vec{r}_i}.
    \label{kin}
\end{equation}
If the system only contains particles with finite mass, the kinetic
energy operator can be rewritten as a sum of the kinetic energy
operators of the relative and center of mass motion and the center of
mass motion can be easily eliminated \cite{suzuki1998stochastic}.
$V$ is the Coulomb interaction
\begin{equation}
    V=\sum_{i<j}^{N} {\frac{q_iq_j}{|\vec{r}_i-\vec{r}_j|}}.
\end{equation}
$U$ is the nuclear potential in the case of fixed (infinite mass) nuclei 
\begin{equation}
    U=\sum_{j=1}^{N_e}\sum_{i=1}^{N_{nuc}} \frac{q_j q_i}{(\vec{r}_j-\vec{R}_i)},
\end{equation}
and the dipole moment $\vec{D}$ of the system is defined as
\begin{equation}
    \vec{D}=\sum_{i=1}^N q_i\vec{r}_i.
\end{equation}

The operators act in real space, except $q$ which acts on the photon space
\begin{eqnarray}
q|n\rangle&=&\frac{1}{\sqrt{2
\omega}}\left(a+a^{+}\right)|n\rangle\\
&=&
\frac{1}{\sqrt{2 \omega}}\left(\sqrt{n}|
n-1\rangle+\sqrt{n+1}|n+1\rangle\right)\nonumber .
\end{eqnarray}
\subsection{Trial functions}
Introducing the shorthand notations  $\vec{r}=(\mathbf{r}_1,...,\mathbf{r}_N)$, and
$\vert n \rangle$ 
where $n$ is  the number of photons in photon mode $\omega$, 
the variational trial wave function is written as a linear
combination of  products of spatial and photon
space basis functions
\begin{equation}
\Psi(\vec{r})=\sum_{{n}}\sum_{k=1}^{K_{{n}}}
    c_k^{{n}}\psi_k^{{n}}(\vec{r})|{n}\rangle.
    \label{pwfM}
\end{equation}

The spatial part of the wave function is expanded into ECGs 
for each photon state $|\vec{n}\rangle$ as
\begin{equation}
    \psi_k^{{n}}(\vec{r})={\cal A}\lbrace {\rm e}^{-{1\over 2}\sum_{i<j}^N
    \alpha_{ij}^k(\mathbf{r}_i-\mathbf{r}_j)^2-{1\over 2}\sum_{i=1}^N 
    \beta_i^k(\mathbf{\mathbf{r}}_i-\mathbf{s}_i^k)^2}
    \Lambda({\vec{r}})
    \chi_S\rbrace
\label{bf}
\end{equation}
where ${\cal A}$ is an antisymmetrizer, $\chi_S$ is the $N$ particle spin function 
(coupling the spin to $S$), and 
$\alpha_{ij}^k$,$\beta_i^k$ and $\mathbf{s}_i^k$ are nonlinear parameters.

The DSI introduces a non-spherical term 
into the Hamiltonian. The solution of this non-spherical problem is very
difficult and slowly converging. 
To avoid this we introduce  $\Lambda(\vec{r})={\rm e}^{\vec{r}U\vec{r}}$
in  Eq. \eqref{bf} to eliminate the DSI term $H_d$
from Eq. \eqref{hep} altogether. In the exponential, $U$ is a $3N\times 3N$ matrix
with elements chosen in such a way that when the kinetic energy acts
on the trial function, the resulting expression cancels the DSI term
\cite{PhysRevLett.127.273601}.

The necessary matrix elements can be analytically computed for both the spatial 
and the photon components \cite{doi:10.1063/5.0066427,
PhysRevLett.127.273601}, and the resulting Hamiltonian and 
overlap matrices are highly sparse.

We will optimize the basis functions by selecting the best spatial basis parameters and 
photon components using the Stochastic Variational  method (SVM) 
\cite{suzuki1998stochastic,PhysRevC.52.2885}. In the SVM approach, 
a large number of candidate basis functions are randomly generated, 
and the ones that yield the lowest energy are chosen
\cite{PhysRevLett.127.273601,suzuki1998stochastic,PhysRevC.52.2885}. 
The basis size can be increased by adding the best states one by one, 
and a K-dimensional basis can be refined by replacing states with 
randomly selected better basis functions. This approach is very efficient 
in finding suitable basis functions.

\section{Results and discussion}
Three systems, the $H^-$ and $H_2^+$ ions and the He atom is used as
example. Atomic units will be used 
($m_e$=1, $\hbar=1$ and $e$=1) and the mass of the
proton and the He nucleus is expressed in in electron mass $m_e$. 
To calculate the binding energies we also have to calculate
the energy of the H atom and He$^+$ ion. The nuclei with infinite mass
are positioned at the origin, except for the H$_2^+$ ion the positions
are $\vec{R}_1=(-d/2,0,0)$ and $\vec{R}_2=(d/2,0,0)$. For the coupling
$\vec{\lambda}=(\lambda,0,0)$ is chosen, and $\lambda$ will be varied
between $\lambda=0.01$ and $\lambda=0.1$. The experimentally achievable
$\lambda$ value is somewhere below $\lambda=0.05$ and most calculations
use the 0.01-0.1 range. The
basis size is 100 for one-particle cases ($N=1$,H atom, H$_2^+$ and He$^+$
ions with infinite nuclear mass) 400 for two-particle cases 
($N=2$, H$^-$, He with infinite mass) and 1000 for the three-particle systems.
The nonlinear parameters are optimized until the energy converged in
the  first 6 decimals.

\subsection{The H atom}
\begin{table}
\begin{tabular}{|l|l|l|l|l|}
\hline
$\lambda$& $E_0$ & $E$        & $p_0$    & $p_1$ \\\hline
0.01 & -0.499675 & -0.499691  & 0.999975 & 2.5$\times 10^{-5}$ \\
0.02 & -0.499521 & -0.499590  & 0.999900 & 1.0$\times 10^{-4}$\\ 
0.03 & -0.499275 & -0.499421  & 0.999776 & 2.2$\times 10^{-4}$\\
0.04 & -0.498925 & -0.499184  & 0.999605 & 3.9$\times 10^{-4}$\\
0.05 & -0.498484 & -0.498883  & 0.999388 & 6.1$\times 10^{-4}$\\
0.06 & -0.497933 & -0.498515  & 0.999120 & 8.7$\times 10^{-4}$\\
0.07 & -0.497308 & -0.498071  & 0.998820 & 1.2$\times 10^{-3}$\\
0.08 & -0.496585 & -0.497579  & 0.998468 & 1.5$\times 10^{-3}$\\
0.09 & -0.495773 & -0.497014  & 0.998081 & 1.9$\times 10^{-3}$\\
0.10 & -0.494873 & -0.496385  & 0.997660 & 2.3$\times 10^{-3}$\\
\hline
\end{tabular}
\caption{Properties of a H atom with finite proton mass 
($m=1836.1515$) and $\omega=0.22$.}
\label{table:H}
\end{table}
Table \ref{table:H} shows the energy of the H atom  as a function of
$\lambda$ for the finite proton mass case. The first energy, $E_0$, 
is the energy of the atom without
coupling to photon spaces, the energy change in this case is purely
due to the DSI term
$\frac{1}{2}(\vec{\lambda}\cdot\vec{D})^2$. The second energy, $E$, in
Table \ref{table:H} is the energy of the system coupled to photon spaces
$\vert n\rangle$, $n\ge 0$. 
We also show the probability of the wave function in the zero photon
space ($p_0$) and the one photon space $(p_1)$. As the DSI
term is positive, the energy of the H atom increases
with $\lambda$ for both $E_0$ and $E$. The probability of the $\vert
1\rangle$ photon space is small but increasing with $\lambda$. The
probabilities of the higher photon spaces (not shown) are typically
10$^{-3}$ times smaller, $p_{n+1}\approx 10^{-3} p_n$. These probabilities also
increase with $\lambda$ and for $\lambda>0.05$ photon spaces up to
$n=6$ contribute to the energy in the 5th or 6th decimals. 

\begin{table}
\begin{tabular}{|l|l|l|l|l|}
\hline
$\lambda$& $E_0$ & $E$        & $p_0$    & $p_1$ \\\hline
0.01 & -0.499932 & -0.499946  & 0.999976 & 2.4$\times 10^{-5}$ \\
0.02 & -0.499789 & -0.499844  & 0.999909 & 9.0$\times 10^{-5}$\\ 
0.03 & -0.499542 & -0.499678  & 0.999793 & 2.1$\times 10^{-4}$\\
0.04 & -0.499189 & -0.499447  & 0.999615 & 3.8$\times 10^{-4}$\\
0.05 & -0.498737 & -0.499132  & 0.999386 & 6.1$\times 10^{-4}$\\
0.06 & -0.498191 & -0.498783  & 0.999126 & 8.7$\times 10^{-4}$\\
0.07 & -0.497569 & -0.498348  & 0.998824 & 1.2$\times 10^{-3}$\\
0.08 & -0.496840 & -0.497850  & 0.998477 & 1.5$\times 10^{-3}$\\
0.09 & -0.496027 & -0.497282  & 0.998095 & 1.9$\times 10^{-3}$\\
0.10 & -0.495115 & -0.496655  & 0.997679 & 2.3$\times 10^{-3}$\\
\hline
\end{tabular}
\caption{Properties of a H atom with infinite proton mass and
$\omega=0.22$.}
\label{table:H_inf}
\end{table}
The infinite mass case (Table \ref{table:H_inf}) shows very similar
tendency, the photon space probabilities barely changed, the energies
are slightly decreased due to the the increased mass.

\subsection{The H$^-$ ion}
\begin{figure}
\includegraphics[width=0.45\textwidth]{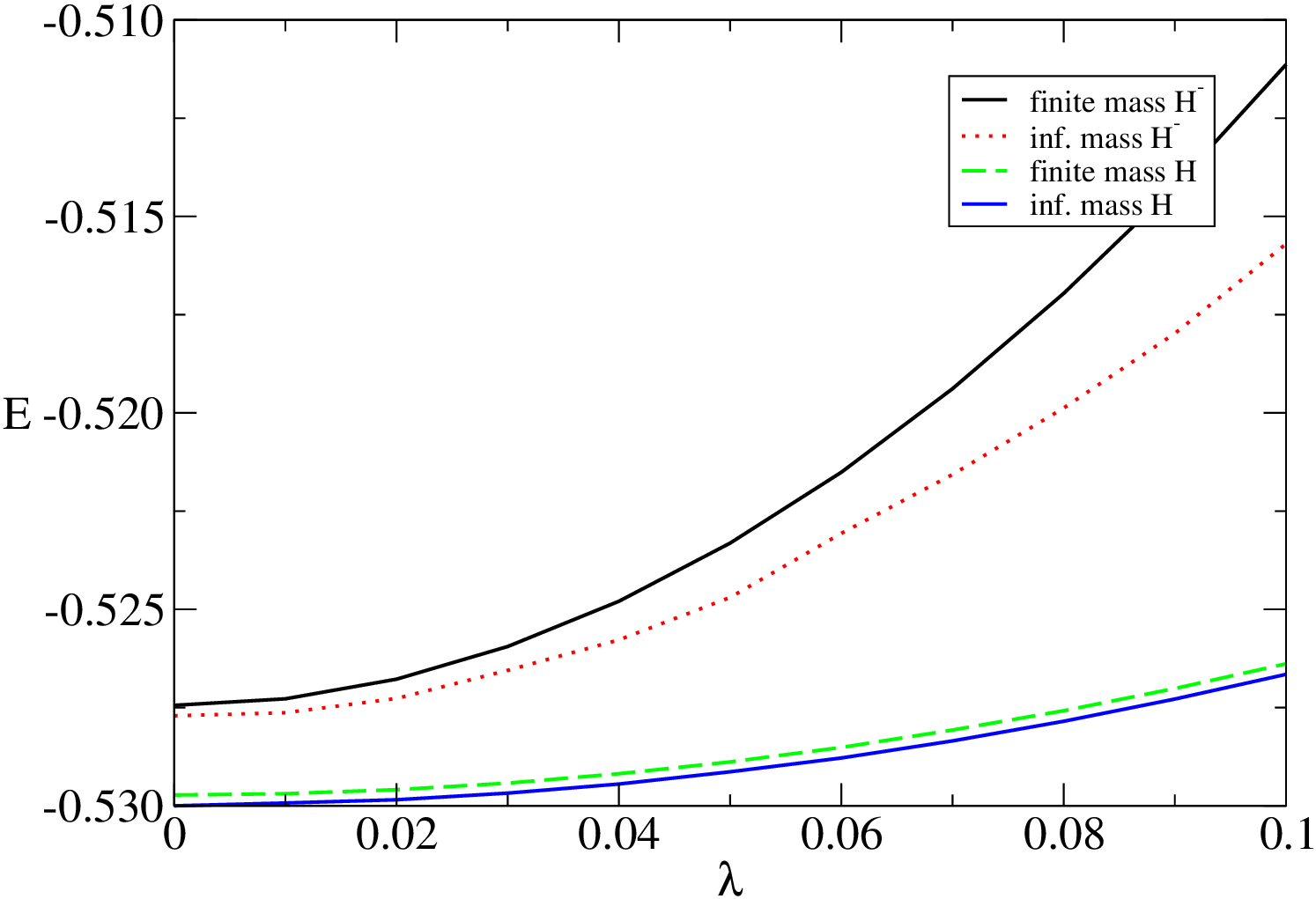}
\caption{Energy of the H atom and the H$^-$ ion as a function of $\lambda$ for
$\omega=0.22$. The energy of the H atom is shifted by -0.03 so that
the two systems can be shown in the same figure.}
\label{fig:H}
\end{figure}

First, we show the calculation for the H$^-$ ion with finite proton mass.
The effect of the cavity is much larger on the H$^-$ ion, as expected
(see Table \ref{table:H-}).
This system is weakly bound, the dipole moment is larger and couples
to the light much more strongly. The DSI ($E_0$
column in Table \ref{table:H-}) strongly increases the energy. The
energy of the light-matter coupled system, $E$, also increases with
$\lambda$ but not as strongly as $E_0$. The coupling changes the energy
in the second decimal and the probability of the zero photon space
decreases to 0.95. The tendency is very similar for the infinite mass
case but the energies are significantly different. 
This is illustrated in Fig. \ref{fig:H}. The energy of the
finite and infinite mass case changes to a different extent in the H$^-$
ion case, while in the case of the H atom the two energies change much
less with $\lambda$ and behave almost identically.

\begin{table}
\begin{tabular}{|l|l|l|l|l|}
\hline
$\lambda$& $E_0$ & $E$        & $p_0$    & $p_1$ \\\hline
0.01 &  -0.527025  & -0.527276 & 0.999152 & $8.4\times 10^{-4}$\\
0.02 &  -0.525849  & -0.526777 & 0.996681 & $3.2\times 10^{-3}$\\
0.03 &  -0.524030  & -0.525946 & 0.992822 & $6.9\times 10^{-3}$\\ 
0.04 &  -0.521667  & -0.524794 & 0.987690 & $1.2\times 10^{-2}$\\
0.05 &  -0.518830  & -0.523313 & 0.981802 & $1.7\times 10^{-2}$\\
0.06 &  -0.515572  & -0.521512 & 0.975452 & $2.2\times 10^{-2}$\\
0.07 &  -0.511936  & -0.519387 & 0.968746 & $2.7\times 10^{-2}$\\
0.08 &  -0.507957  & -0.516961 & 0.962100 & $3.2\times 10^{-2}$\\
0.09 &  -0.503653  & -0.514233 & 0.955476 & $3.7\times 10^{-2}$\\
0.10 &  -0.499061  & -0.511135 & 0.950017 & $4.1\times 10^{-2}$\\
\hline
\end{tabular}
\caption{Properties of a H$^-$ ion with finite proton mass
($m=1836.1515$) and
$\omega=0.22$.}
\label{table:H-}
\end{table}

\begin{table}
\begin{tabular}{|l|l|l|l|l|}
\hline
$\lambda$& $E_0$ & $E$        & $p_0$    & $p_1$ \\\hline
0.01 & -0.527377 & -0.527632 &  0.999159 & $8.4\times 10^{-4}$\\
0.02 & -0.526318 & -0.527268 &  0.996756 & $3.2\times 10^{-3}$\\
0.03 & -0.524660 & -0.526551 &  0.993994 & $5.9\times 10^{-3}$\\
0.04 & -0.522501 & -0.525778 &  0.988958 & $1.1\times 10^{-2}$\\
0.05 & -0.520096 & -0.524732 &  0.983072 & $1.6\times 10^{-2}$\\
0.06 & -0.517207 & -0.523379 &  0.980888 & $1.8\times 10^{-2}$\\
0.07 & -0.514182 & -0.521570 &  0.975456 & $2.3\times 10^{-2}$\\
0.08 & -0.510917 & -0.519876 &  0.966552 & $3.0\times 10^{-2}$\\
0.09 & -0.507437 & -0.517978 &  0.961433 & $3.3\times 10^{-2}$\\
0.10 & -0.503754 & -0.515700 &  0.956274 & $3.7\times 10^{-2}$\\
\hline
\end{tabular}
\caption{Properties of a H$^-$ ion with infinite proton mass and
$\omega=0.22$.}
\end{table}

\begin{figure}
\includegraphics[width=0.45\textwidth]{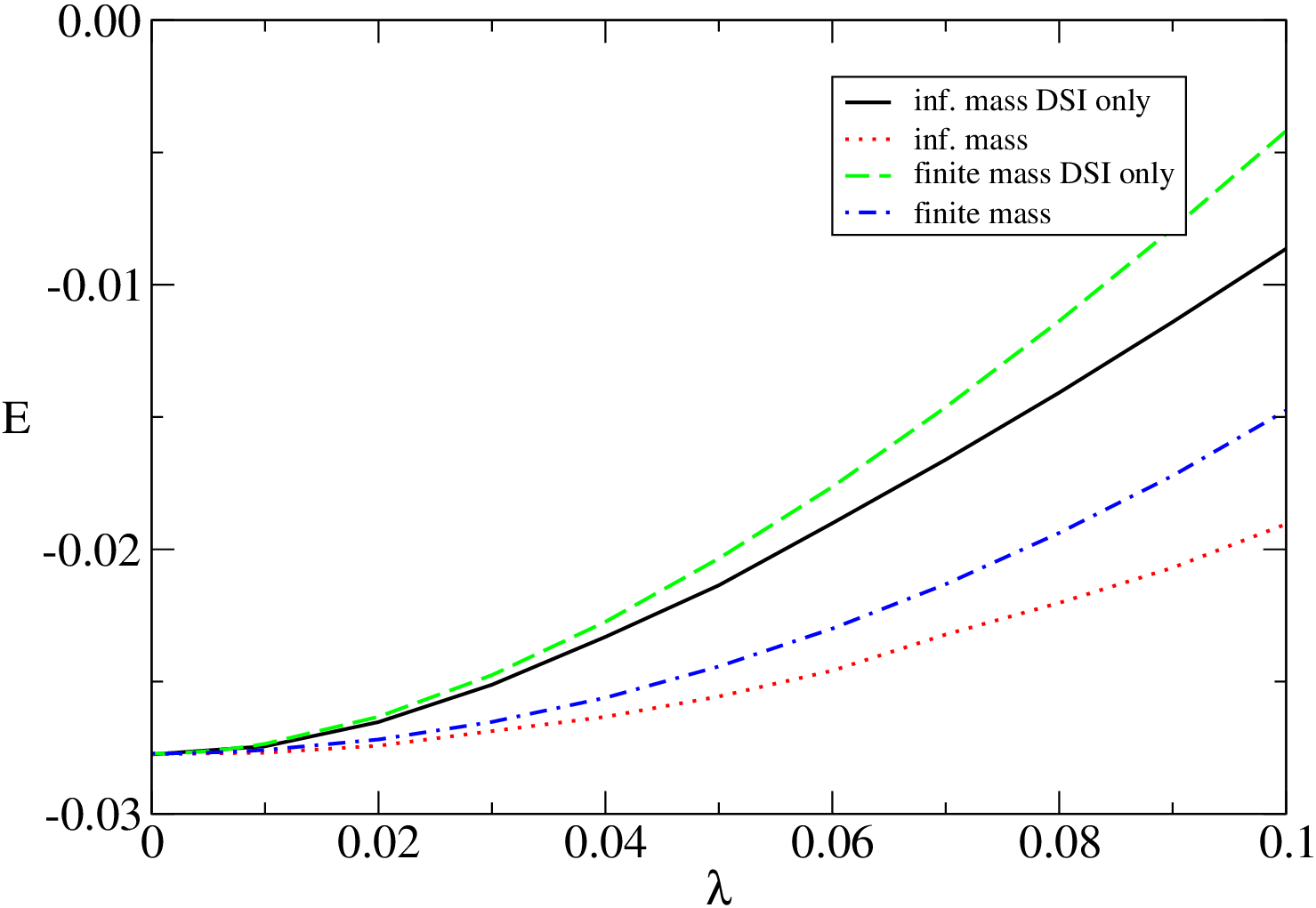}
\caption{Binding energy of the H$^-$ ion as a function of $\lambda$ for
$\omega=0.22$.}
\label{fig:H-}
\end{figure}

Fig. \ref{fig:H-} shows the binding energy of the H$^-$ ion in the
finite and infinite proton mass cases. The binding energy decreases
as  $\lambda$ increases and the binding energy of the finite mass case 
decreases faster then the infinite one. The binding energy change due
to the DSI alone behave similarly and the difference between the
difference between the DSI total binding energy curves show the
importance of the coupling to the light spaces.


\subsection{The He atom}
\begin{figure}
\includegraphics[width=0.45\textwidth]{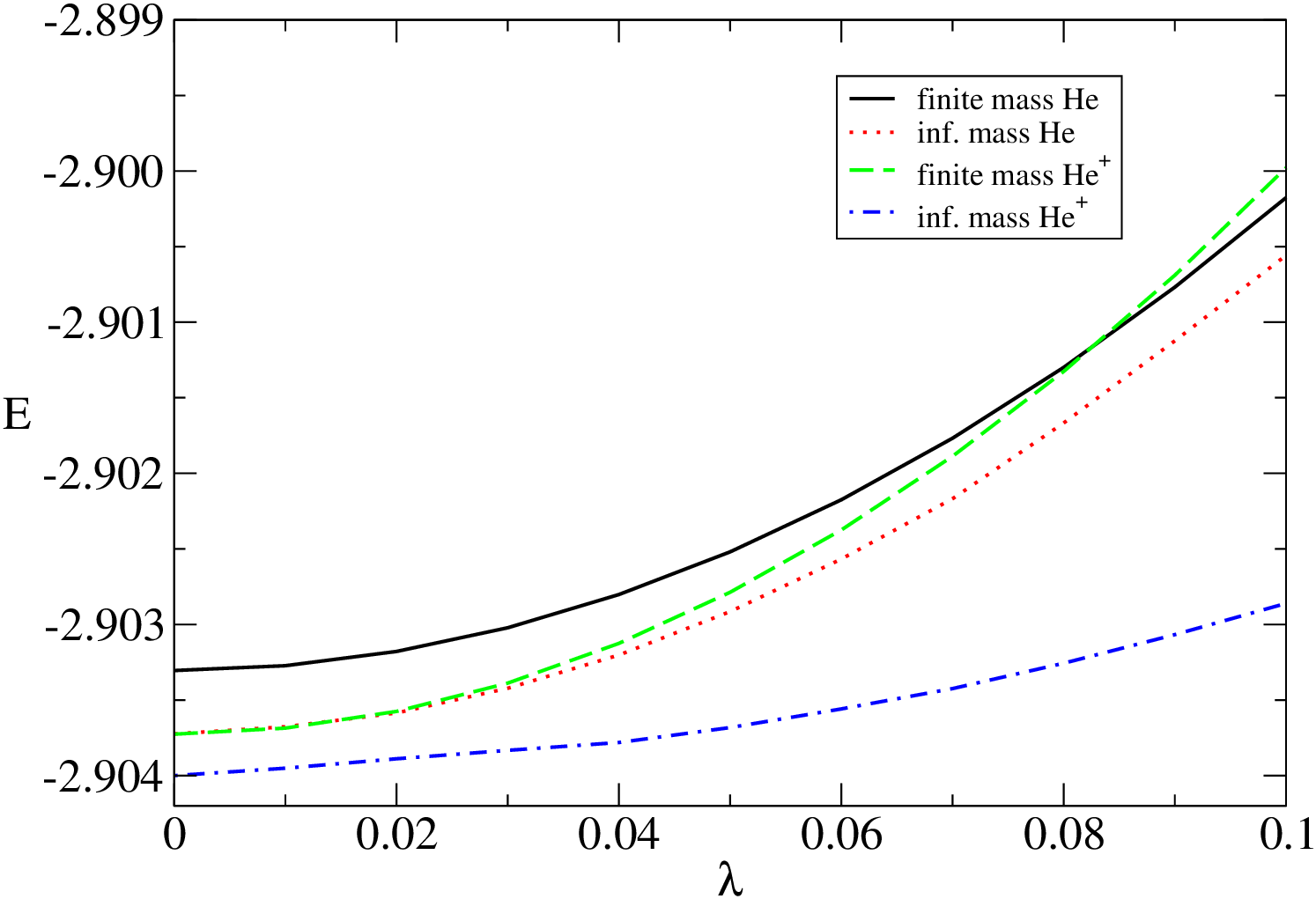}
\caption{Energy of the He atom and the He$^+$ ion as a function of $\lambda$ for
$\omega=0.22$. The energy of the He$^+$ ion is shifted by -0.904.}
\label{fig:He}
\end{figure}

\begin{figure}
\includegraphics[width=0.45\textwidth]{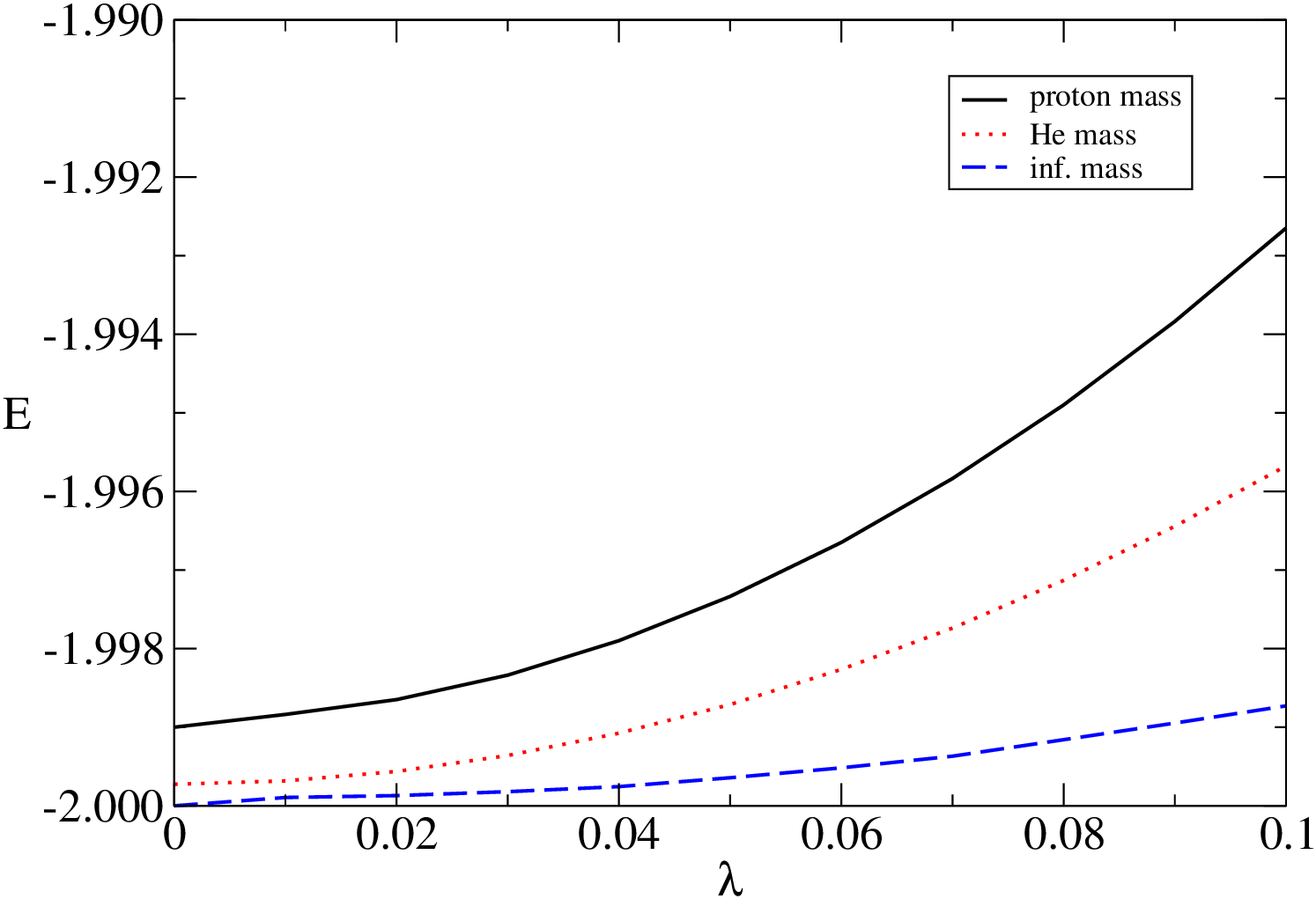}
\caption{Mass dependence of the energy of the He$^+$ ion as a function of $\lambda$ for
$\omega=0.22$.}
\label{fig:mass}
\end{figure}

\begin{figure}
\includegraphics[width=0.45\textwidth]{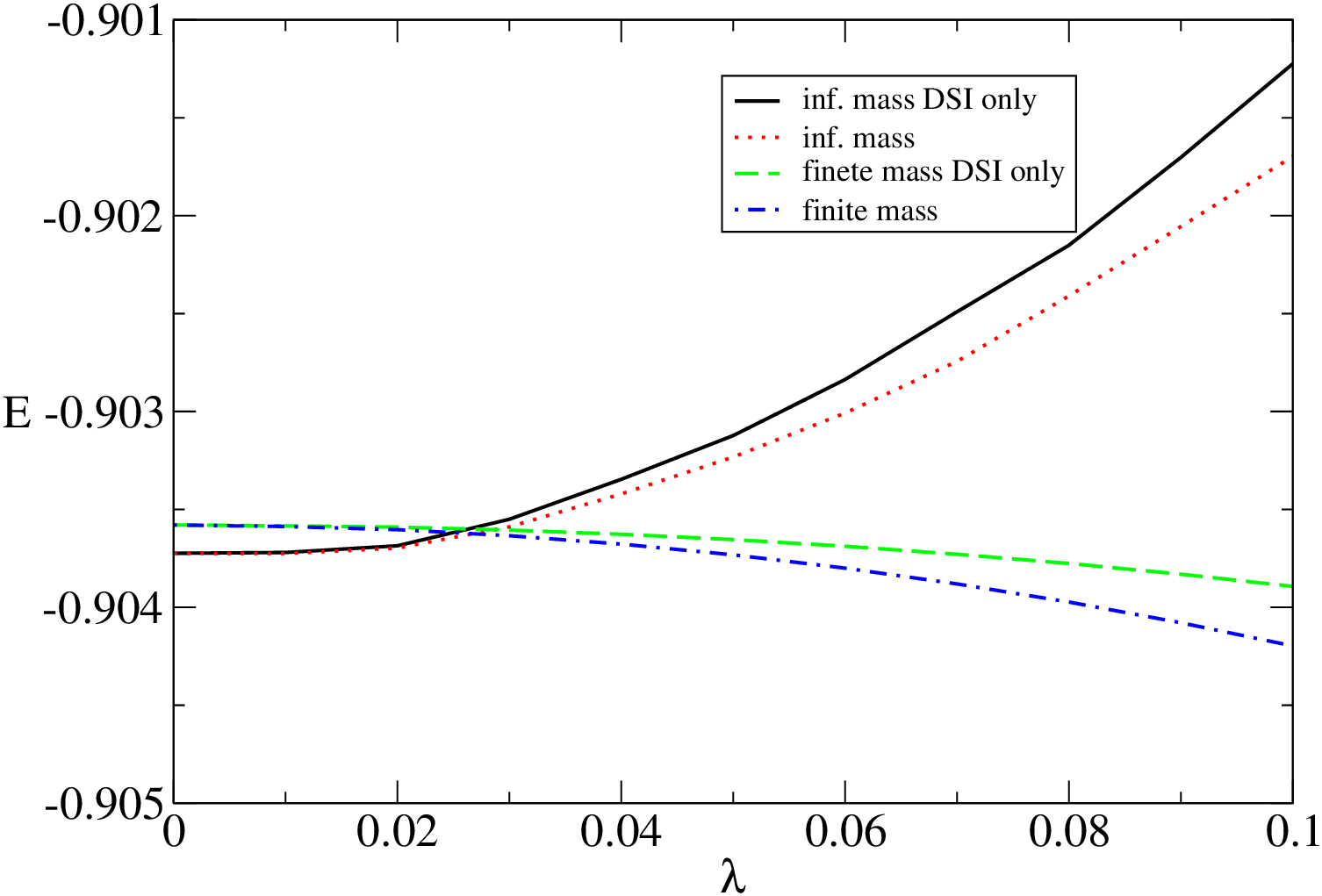}
\caption{Binding energy of the He atom as a function of $\lambda$ for
$\omega=0.22$.}
\label{fig:He_bin}
\end{figure}

The energies of the He atom with finite and infinite nuclear mass are
listed in Tables \ref{table:He} and \ref{table:He_inf}. Compared to
the H$^-$ ion the electrons are strongly bound in the He atom and the
energy change is much less when $\lambda$ is increased. The photon
space probabilities also remain very low about 10$^{-4}$. The energy
of the He atom and the He$^+$ ion is also shown as the function of
$\lambda$ in Fig. \ref{fig:He}. The energy curves of the finite and
infinite mass He atom are very similar, but the $\lambda$ dependence 
of the energy of finite and infinite mass He$^+$ are significantly
different. This is further investigated in Fig. \ref{fig:mass}, where 
we add a calculation using a smaller artificial mass (taken to be equal
to the mass of a proton). Fig. \ref{fig:mass} shows that the energy is
increasing faster as the function of $\lambda$ for lighter particles. 
This leads to a very interesting case for the binding energies shown
in Fig. \ref{fig:He_bin}. The binding energy decreases with increasing 
$\lambda$ for the infinite mass case, while the  binding energy
increases with increasing $\lambda$ for the finite mass case. This is 
true for both the SDI and the full coupled Hamiltonian. As the binding 
energy of the infinite mass case is larger than the finite mass case
there is a crossover at around $\lambda=0.025$. 
\begin{table}
\begin{tabular}{|l|l|l|l|l|}
\hline
$\lambda$& $E_0$ & $E$         & $p_0$    & $p_1$ \\\hline
0.00 & -2.903305 & -2.903305   &   1.0    &   0.0\\
0.01 & -2.903267 & -2.903273   & 0.999995 & 5.0$\times 10^{-6}$\\
0.02 & -2.903153 & -2.903179   & 0.999980 & 2.0$\times 10^{-5}$\\
0.03 & -2.902966 & -2.903022   & 0.999954 & 4.6$\times 10^{-5}$\\
0.04 & -2.902703 & -2.902802   & 0.999919 & 8.1$\times 10^{-5}$\\
0.05 & -2.902365 & -2.902510   & 0.999874 & 1.3$\times 10^{-4}$\\
0.06 & -2.901953 & -2.902175   & 0.999820 & 1.8$\times 10^{-4}$\\
0.07 & -2.901467 & -2.901768   & 0.999755 & 2.4$\times 10^{-4}$\\
0.08 & -2.900907 & -2.901299   & 0.999685 & 3.2$\times 10^{-4}$\\
0.09 & -2.900274 & -2.900768   & 0.999606 & 3.9$\times 10^{-4}$\\
0.10 & -2.899569 & -2.900176   & 0.999527 & 4.7$\times 10^{-4}$\\
\hline
\end{tabular}
\caption{Properties of a He atom with finite  mass (7294.26) and
$\omega=0.22$.}
\label{table:He}
\end{table}

\begin{table}
\begin{tabular}{|l|l|l|l|l|}
\hline
$\lambda$& $E_0$ & $E$        & $p_0$    & $p_1$ \\\hline
0.00  & -2.903724  & -2.903724    & 1.0      & 0.\\
0.01  & -2.903669  & -2.903677    & 0.999997 & 3.1$\times 10^{-6}$\\
0.02  & -2.903556  & -2.903586    & 0.999984 & 1.6$\times 10^{-5}$\\
0.03  & -2.903370  & -2.903422    & 0.999961 & 3.9$\times 10^{-5}$\\
0.04  & -2.903102  & -2.903202    & 0.999924 & 7.5$\times 10^{-5}$\\
0.05  & -2.902766  & -2.902915    & 0.999883 & 1.2$\times 10^{-4}$\\
0.06  & -2.902354  & -2.902567    & 0.999834 & 1.7$\times 10^{-4}$\\
0.07  & -2.901860  & -2.902168    & 0.999759 & 2.4$\times 10^{-4}$\\
0.08  & -2.901310  & -2.901667    & 0.999710 & 2.9$\times 10^{-4}$\\
0.09  & -2.900649  & -2.901123    & 0.999647 & 3.5$\times 10^{-4}$\\
0.10  & -2.899953  & -2.900558    & 0.999512 & 4.8$\times 10^{-4}$\\
\hline
\end{tabular}
\caption{Properties of a He atom with infinite  mass,
$\omega=0.22$.}
\label{table:He_inf}
\end{table}

\begin{figure}
\includegraphics[width=0.45\textwidth]{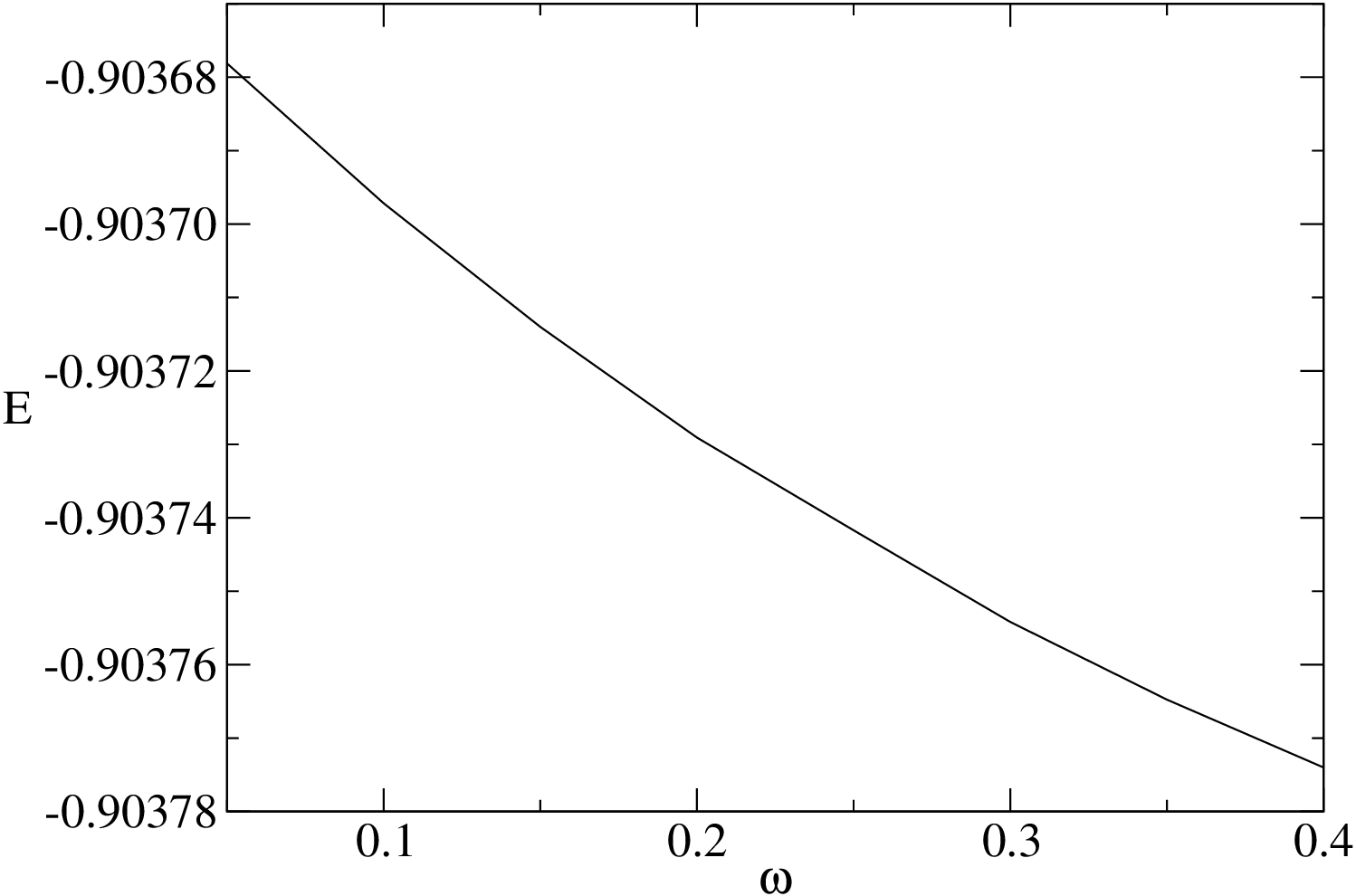}
\caption{Binding energy of the He atom as a function of  $\omega$ for
$\lambda=0.05$ with finite nuclear mass.}
\label{fig:omega}
\end{figure}

Finally, a note about the dependence on the cavity frequency. We have
used $\omega=0.22$ in the calculations so far, but the results would
barely change for different $\omega$. As shown in Fig. \ref{fig:omega}
the change in the energy is very small for a wide range of $\omega$,
the binding energy only changes in the fifth digit.


\section{The H$_2^+$ ion}
\begin{figure}
\includegraphics[width=0.45\textwidth]{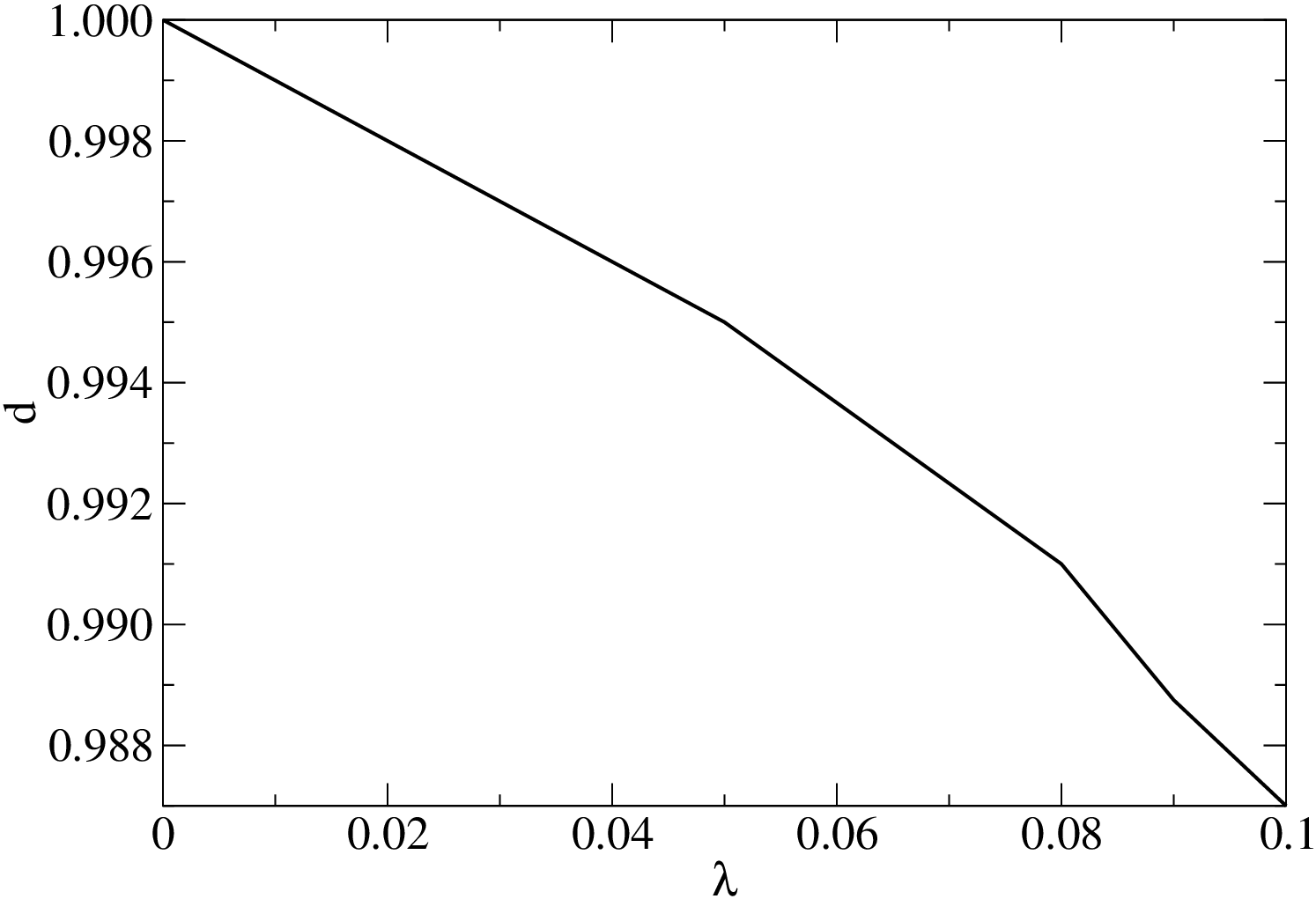}
\caption{Bond length of H$_2^+$ ion as a function of  $\lambda$ for
$\omega=0.1$.}
\label{fig:bond}
\end{figure}

\begin{figure}
\includegraphics[width=0.45\textwidth]{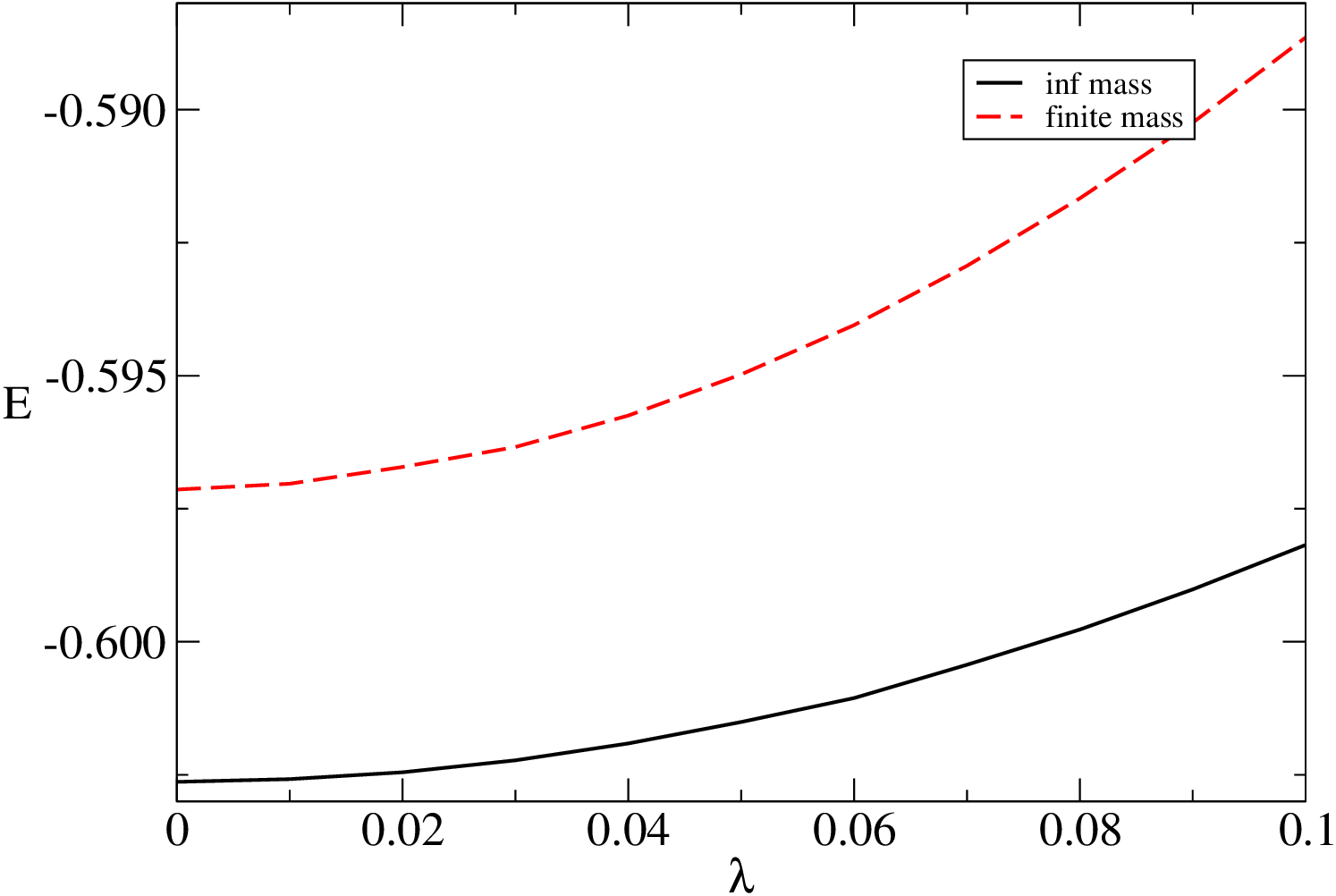}
\caption{Energy of the H$_2^+$ ion as a function of $\lambda$ for
$\omega=0.1$.}
\label{fig:H2+}
\end{figure}

\begin{figure}
\includegraphics[width=0.45\textwidth]{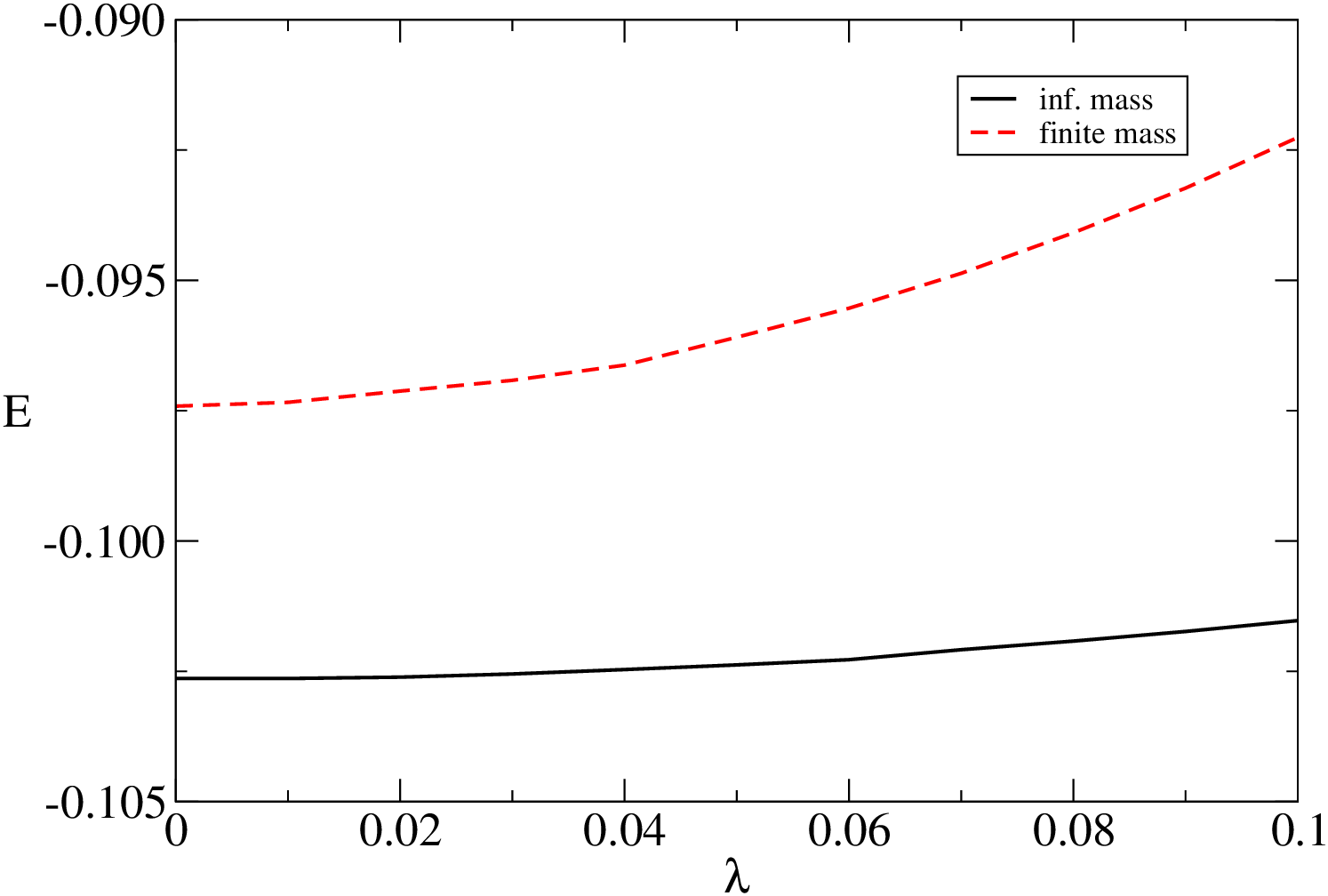}
\caption{Binding energy of the H$_2^+$ ion as a function of $\lambda$ for
$\omega=0.1$.}
\label{fig:H2+_bin}
\end{figure}

The final example is the H$_2^+$ molecular ion. In this case, for the
infinite mass case we first calculated the equilibrium bond length as
the function of $\lambda$ and then fixed the distance between the two
protons at the equilibrium and calculated the energy of the ion. As
shown in Fig. \ref{fig:bond} the bond length gets slightly lower with
increasing $\lambda$.  Figs. \ref{fig:H2+} and \ref{fig:H2+_bin} shows
the total energies and the binding energies of the finite and infinite 
mass cases as a function of $\lambda$. In this case, the energies
behave very similarly except for an overall shift (the infinite mass
case have larger total and binding energy), and the energy of the
finite mass case changes somewhat more.


\section{Summary}
The effect of finite nuclear mass is investigated in coupled
light-matter systems in cavity QED using the
Pauli-Fierz Hamiltonian. Three different systems are investigated: the
helium atom, the hydrogen negative ion (H$^-$), and the hydrogen
molecular ion (H$_2^+$). The study finds small but significant differences
in the behavior of the binding energies as a function of the coupling
strength. The binding energy decreases for H$_2^+$ and H$^-$ as
$\lambda$ increases for both finite and infinite mass in a similar way.
For the He atom, however, the binding energy decreases for infinite mass
and increases for finite mass. These differences are due to the
competition of the kinetic energy terms and the dipole moment
depending parts of the Hamiltonian. In the infinite mass case the
nuclei are fixed and the nuclear coordinates do not contribute to the
total dipole of the system. In the finite nuclear mass case the
nuclear motion also effects the dipole moment and there is a balance
between the kinetic energy of the nuclei and the nuclear dipole
dependence of the PF Hamiltonian.

Additionally, the probability of coupling to light is found
to be very small, but even this small coupling has a strong effect on
the energies of the systems. For tightly bound systems like the H or 
He atoms 99\% of the wave function is in the zero photon space for
realistic values of $\lambda$. For weakly bound systems this probability
drops to 95\% and higher photons spaces become important as well.

These very accurate test calculations can serve as benchmark cases of
QEDFT, QED-CC \cite{doi:10.1063/5.0078795,doi:10.1063/5.0038748,PhysRevX.10.041043,
PhysRevResearch.2.023262} or configuration interaction
\cite{doi:10.1021/acs.jctc.3c01207,10.1063/5.0230565,matousek_polaritonic_2024}
calculations. These calculations also show that the nuclear motion can be
very important in cavity QED calculations. The infinite mass
approximation can be corrected by the Born-Oppenheimer expansion in
molecular calculations. Similar approach has been worked out for the
cavity QED case  \cite{PhysRevA.98.043801}. The solution of the
coupled nucleus-electron-photon case, however, is complicated.

\begin{acknowledgments} 
This work has been supported by the National Science
Foundation (NSF) under Grant No. IRES 2245029 and DMR-2217759.
\end{acknowledgments}

\section*{Data Availability Statement}
The data that support the findings of this study are available
from the corresponding author upon reasonable request.

\section*{AUTHOR DECLARATIONS}
\par\noindent
{\bf Conflict of Interest}

The authors have no conflict of interest to disclose.

%

\end{document}